\documentstyle[aps,amsfonts,amssymb,twocolumn]{revtex}
\begin{document}
\draft

\title{Heavy Particle in a d-dimensional Fermionic Bath:
A Strong Coupling Approach}
\author{Achim Rosch and Thilo Kopp}
\address{Institut f\"ur Theorie der Kondensierten Materie, Universit\"at
Karlsruhe,
D--76128 Karlsruhe, Germany}

\date{February 8, 1995}
\maketitle

\begin{abstract}
We investigate the low temperature behavior of a heavy particle (mass $M$)
in a fermionic bath. An effective action is considered which exactly implements
the orthogonality catastrophe. It is equivalent to a model of
local bosons coupling via a long-range interaction to the particle. The saddle
point solution for the heavy particle propagator yields a finite quasiparticle
weight
in two and higher dimensions. In one dimension the propagator decays
algebraically
for long times. The saddle point approximation is valid for large $M$.
\end{abstract}


The low temperature ($T\!\rightarrow\!0$) behavior of a heavy particle strongly
coupled
to a fermionic bath has been a fascinating puzzle for more than two decades and
yet has
not been solved. Specifically, the question has been raised if, for
$T\!\rightarrow\! 0$,
the particle has a plane  wave character, or if the particle becomes a totally
incoherent
excitation so that the quasi-particle (QP) picture fails. The low temperature
mobility of the heavy particle and the existence of a finite Drude weight are
related
problems.

A variety of real physical systems are assumed to be
`heavy-particle-fermionic-bath' systems:
e.g.\ ions in normal liquid $^3$He \cite{josephson,woelfle,kondo},
diffusion of myons and hydrogen in metals \cite{kagan}, photoemission spectra
of
metals \cite{doniach}, and valence band holes in n-type doped
semiconductors \cite{calleja}. Apart from these applications,
a genuine theoretical interest in the solution of such a principal problem
exists:
it is closely related to the Kondo problem and the dissipative 2-level system
\cite{leggett}.
Further, it may give insight into the possible breakdown of Fermi liquid theory
for strongly
correlated electron systems: a single overturned spin in an otherwise
spin-polarized
Fermi gas with local interactions (Hubbard model) may or may not display a QP
behavior \cite{mcguire,kopp,frenkel,castella,sorella}. The existence of a
QP-peak in the
spectral function of the overturned spin
will depend on dimension---and possibly on its effective mass and interaction
with the
Fermi bath. Of course, the (non-) existence of a QP-peak in the single-spin
flip case does
not predecide about the (non-) Fermi liquid behavior for the paramagnetic case.
But it may
be assumed that a Fermi sea of flipped spins will rather stabilize the Fermi
liquid
behavior because of the restricted phase space available for scattering
processes. Also,
the overturned spin is not necessarily a heavy particle. However the case of a
heavy
particle should be investigated first because it is the most probable to reveal
non-QP
behavior,  as the infinitely heavy particle does.

In this paper we will focus on the question if the particle shows coherent
behavior
for $T\!\rightarrow\!0$, i.e.\ if its spectral function displays a
$\delta$-function peak for
zero momentum. The physics of this low-energy  sector of the system is
determined by two competing phenomena: orthogonality catastrophe (OC)
\cite{anderson}
versus recoil \cite{doniach,muellerh}. On the one hand, we expect an incoherent
behavior because of Anderson's
OC valid for an infinitely heavy particle: a sudden displacement of the
particle
will produce an infinite number of low-energy particle-hole (ph) excitations.
Similarly, a localized valence band hole generated
through optical absorption at time $0$ will produce a shakeup of the conduction
band
in the long time limit.
This causes a vanishing overlap
of the original  state with the final state in the thermodynamic
limit. The final state is characterized by finite phase shifts of all the bath
particles,
necessary to screen the localized hole. On the other
hand, momentum conservation for a finite-mass particle
implies a finite recoil energy up to $E_{\text{recoil}} \simeq (2k_f)^2/2M$
of the heavy particle  with mass $M$. This finite $E_{\text{recoil}}$ restricts
the available phase space for scattering processes on the Fermi surface simply
by energy conservation. Accordingly, a perturbative calculation in the
interaction
of the heavy particle with the fermionic bath shows that the number of low
lying ph-excitations is reduced in space dimensions $d>1$ so substantially that
the QP-peak is restored for finite mass \cite{kopp}.
However, it has been argued that conclusive evidence for a finite QP-peak in
$2d$
has to be obtained from a non-perturbative approach---even if the interaction
is small.

To set up the problem  we consider the following Hamiltonian:
\begin{equation}
H=H_{o} + H_{\text{bath}} + H_{U}
\label{H}\end{equation}
Here, the kinetic energy of the heavy particle with momentum $\bbox{P}$ is
$$H_{o}=\bbox{P}^{\,2}/(2M)$$
For some specific lattice cases, it would be more appropriate to
assume a band dispersion. We will comment on such a case below.
Actually, it is important to include a periodic potential for the discussion of
self-trapping \cite{schmid,zwerger,sols}. But it has been asserted that,
even for charge $Z\ge 2$, localization is unlikely to occur \cite{fisher}.

The bath contribution is just a Fermi (lattice) gas with dispersion
$\epsilon_{\bbox{k}}$
$$H_{\text{bath}}=\sum\nolimits_{\bbox{k}} \epsilon_{\bbox{k}}
         c^{\dag}_{\bbox{k}} c_{\bbox{k}}$$
Spin may be included, and the results have to be modified in a trivial way by a
spin degeneracy factor.
The interaction is assumed to be local, i.e.\ it acts at the position
$\bbox{R}$
of the heavy particle:
$$H_{U} =  U \sum\nolimits_{\bbox{kk'}} e^{i\,(\bbox{k'}-\bbox{k})\bbox{R}}
    \; c^{\dag}_{\bbox{k}} c_{\bbox{k'}}$$
Higher scattering channels beyond s-wave should be considered for an `extended
particle'
but here we will investigate only s-wave scattering to keep the mathematics
transparent.

We will present a path integral approach which correctly reproduces the known
limits
(in any dimension): the  second order perturbation theory in
$U$\cite{cumulant},
the high temperature case ($E_{\text{recoil}} < T \ll \epsilon_f$), and the
localized
case ($M=\infty$) which straightforwardly results from the high temperature
case
(with $E_{\text{recoil}} =0$). As regards the coherence of the heavy particle
state,
we find: algebraic decay of the heavy particle propagator in 1$d$,
but finite QP-weight for higher dimensions.  Hence, the finite QP-weight in
$2d$ is
the major result of our considerations.

The partition function for (\ref{H}) is $Z\!=\!Z_o\int\!{\cal D}\bbox{R}\,
e^{-iS[\bbox{R}]}$.
$Z_o$ is the partition function for the fermionic bath. $S\!=\!S_o\!+\!S_{U}$
splits
into the action for the free particle
$S_o\!=\!{1\over2}\int_0^\beta\!d\tau M \dot {\bbox{R}}(\tau)^2$ and the
coupling to
the bath
\begin{equation}
S_{U}=\text{tr}\ln\bigl[\,\openone+
   U\,g_{\scriptscriptstyle\bbox{k}}(\tau\!\!-\!\!\tau')
  e^{i(\bbox{k}-\bbox{k'})\bbox{R}(\tau')}\bigr]
\label{SU}\end{equation}
which results from the integration over the fermionic fields.
$g_{\scriptscriptstyle\bbox{k}}(\tau)$ is  a free-fermion propagator.
The argument of the logarithm is a matrix indexed by $\tau$, $\tau'$,
$\bbox{k}$ and $\bbox{k'}$.
Its power series will generate terms involving any
number of time integrals. In the following paragraphs we will argue that
(\ref{SU}) may
be cast into a `singly retarded' form which requires two time integrations,
only.

A path integral which builds on all the (infrared) singular contributions from
paths
connecting two heavy particle positions was, to our knowledge, first
proposed by Sols and Guinea \cite{sols}. Although we have to refer the reader
to
Prokof'ev's papers \cite{prokof} for the detailed discussion of the set-up of
the action, we
want to highlight the connection to the orthogonality catastrophe (OC).
The OC not only arises for the overlap of a state with undisturbed Fermi sea
and a
state with all particles scattering from a localized impurity.
But also the overlap of the electronic ground state
$\phi_{\bbox{R}}$ with impurity at position $\bbox{R}$ and a state
$\phi_{\bbox{R'}}$
vanishes in the thermodynamic limit $N_e\rightarrow\infty$. $N_e$ is
the number of electrons (bath particles). Yamada and Yoshida et al.\
\cite{yamada}
managed to crack this
integral:
\begin{equation}
\langle\phi_{\bbox{R'}}|\phi_{\bbox{R}}\rangle\rightarrow
  e^{-F(|\bbox{R'}-\bbox{R}|) \log N_e}\label{overlap}
\label{OC}\end{equation}
where
\begin{equation}F(|\bbox{R}|)={1\over\pi^2}\arcsin^2(\sqrt{1-x^2} \sin\delta)
\label{F}\end{equation}
for pure s-wave scattering with phase shift $\delta$, and
 $x\!=\!\langle e^{i\bbox{kR}}\rangle_{|\bbox{k}|=k_f}$ which is $\cos(k_f R)$,
$J_o(k_f R)$ and $\sin(k_f R)/(k_f R)$ for $d\!=\!1$, 2 and 3, respectively.
$F(\bbox{R})$ is also known for extended scatterers with phase shifts in higher
order angular momentum channels \cite{vladar}.
Now we consider the general form of a singly retarded \mbox{(self-)}
interaction.
Due to translational invariance it is certainly a functional of
$\bbox{R}(\tau')\!-\!\bbox{R}(\tau)$, and also of $\tau'\!-\!\tau$. We are only
interested in the low frequency behavior which restricts the scattering
processes
to the vicinity of the Fermi surface where energy and momentum are independent
variables. Accordingly, the retarded interaction separates into a space and
time
dependent part \cite{sols,prokof}:
\begin{equation}
S_{\text{eff}}=S_o  - {1\over2}\int\!\!\!
\int^\beta_0 \!\!\! d\tau d\tau'{\cal F}(|\bbox{R'}\!\!-\!\!\bbox{R}|)\,
\rho(\tau'\!\!-\!\tau)
\label{Seff}\end{equation}
$\rho$ is written as
$$\rho(\tau\!-\tau') ={\pi\over\beta}\sum_{n\neq 0} |\omega_n| \,
  e^{i\omega_n (\tau-\tau')}$$
which reflects the fact that the number of excitations close to
the Fermi surface is proportional to $\omega$, and corrections in
$\omega^2$  are irrelevant for the investigated infrared divergence.
Evaluation of the overlap (\ref{OC}) and comparison with (\ref{F})
fixes $\cal F$: $${\cal F}(R)= ({\delta\over\pi})^2 -F(|\bbox{R}|)$$
which is displayed in the figure. It  includes all paths responsible
for the OC  Eq. (\ref{OC}) \cite{prokof}.

The effective action (\ref{Seff}) may be verified for high temperatures or for
small interaction $U$ by expanding (\ref{SU}) either in
$\bbox{R}\!-\!\bbox{R'}$\cite{highT} or in $U$.
The second order in either expansion is consistent with this singly retarded
action $S_{\text{eff}}$ which, moreover, covers contributions from
any order.

There have been  attempts in the past to extract information about the low-$T$
behavior from this action \cite{sols}. Either it is assumed that the quadratic
terms in $R$ already determine qualitatively the low-$T$ behavior
\cite{prokof}.
Then the system is actually equivalent to a harmonic oscillator bath  with
spectral density
$\sim\omega$ coupled to the heavy particle (Caldeira-Leggett model
\cite{leggett}).
Such a mapping onto a bosonic bath was shown
to be correct for the X-ray edge problem with a localized particle
\cite{schotte}.
Another approach mimics the shape of ${\cal F}(R)$ for $d\geq2$ by assuming
a step-like functional form with the step at $R$ equal the lattice constant
\cite{rosch}.
The latter approach is equivalent to a
procedure by Hedeg{\aa}rd and Caldeira \cite{hedegard} which introduces
independent oscillator baths at each lattice site. They further assumed that if
the particle comes back to a lattice site twice or more times it always
encounters a ``new'' bosonic bath in its thermal equilibrium, i.e.\ the history
of excitations at a site is not
recorded (`non-interacting blip approximation'). Again, as for the harmonic
oscillator bath, no QP-peak shows up for $T\rightarrow0$.

We will demonstrate that the details of the long-distance behavior of ${\cal
F}(R)$,
i.e.\ the Friedel-oscillations, actually control the spectral function of the
heavy particle. Since a direct integration over the field $R$ is hopelessly
complicated we propose the following detour:
to disentangle the untractable retarded part of the action we introduce new
bosonic
fields $b_{\gamma\bbox{k}}(\tau)$. Then the problem is equivalent to the
following polaron-type model, as can be checked by integrating out the bosons:
\begin{equation}
H_b=H_o+\sum_{\gamma\bbox{k}} E_{\gamma}\,
      b^{\dag}_{\gamma\bbox{k}} b_{\gamma\bbox{k}}
     + \sum_{\gamma\bbox{k}} (V_{\bbox{k}}\, b^{\dag}_{\gamma\bbox{k}}\,
     e^{i\bbox{kR}} +\text{c.c.})
\label{Hb}\end{equation}
with a spectral density
$\sum_\gamma \delta(\omega\!-\!E_\gamma)\! =\! \omega\,\Theta(\omega)$
for the local bosons,
and a long-range interaction
$|V_{\bbox{k}}|^2\!=\!\sum_{\bbox{R}}{\cal F}(R) e^{i\bbox{kR}}$.
The particle coordinate is eliminated by a transformation \cite{lee}
$T_{\bbox{R}}\!=\!\exp[i(\sum_{\gamma\bbox{k}}
  \bbox{k}\, b^{\dag}_{\gamma\bbox{k}} b_{\gamma\bbox{k}})\,\bbox{R}\,]$
on the local coordinate system of the heavy particle. The transformed
Hamiltonian reads
$$
H_{\bbox{P}_{\!\!o}}=H_{\text{kin}}({\bbox{P}_{\!\!o}})
   + \sum_{\gamma\bbox{k}}
     E_{\gamma}\, b^{\dag}_{\gamma\bbox{k}}  b_{\gamma\bbox{k}}
   + \sum_{\gamma\bbox{k}} (V_{\bbox{k}}\, b^{\dag}_{\gamma\bbox{k}}
+\text{c.c.})
$$
with
$$H_{\text{kin}}({\bbox{P}_{\!\!o}})=
(\bbox{P}_{\!\!o}-\!\sum_{\gamma\bbox{k}}
                  \bbox{k}\, b^{\dag}_{\gamma\bbox{k}}  b_{\gamma\bbox{k}})^2
/(2M)
$$
where the c-number $\bbox{P}_{\!\!o}$ is the total momentum of the ``polaron''.
With this transformation, $V_{\bbox{k}}$ is just a displacement of
the harmonic oscillators whereas the kinetic energy of the heavy particle
introduces a complicated current-current interaction, the origin of which is
the finite recoil.

We are interested in the real-time heavy particle propagator (at $T=0$):
$$iG_{\bbox{P}_{\!\!o}}(t)=
   \langle0|\langle\bbox{P}_{\!\!o}|e^{-iHt}|\bbox{P}_{\!\!o}\rangle|0\rangle
   \simeq\!\!\int\! {\cal D}[b^\star\!,\!b\,] e^{-iS_{\bbox{P}_{\!\!o}}
[b^\star\!,b;\,t]}$$
$|0\rangle$ is the Fermi sea which then translates to the vacuum for the
bosons of (\ref{Hb}), and  $S_{\bbox{P}_{\!\!o}}$ is the action corresponding
to $H_{\bbox{P}_{\!\!o}}$:
$$
S_{\bbox{P}_{\!\!o}}=S_o({\bbox{P}_{\!\!o}})
  +\sum_{\gamma\bbox{k}} \int_0^t \!dt
 (V_{\bbox{k}}\, b^{\star}_{\gamma\bbox{k}}  +V^\star_{\bbox{k}}\,
b_{\gamma\bbox{k}})
   + S_{\text{int}}
$$
with
$$S_o({\bbox{P}_{\!\!o}})=
    \sum_{\gamma\bbox{k}} \int_0^t \!dt\,
         b_{\gamma\bbox{k}}^\star (
                       -i\partial_t + E_\gamma +{\bbox{k}^2\over 2M} -
                              {\bbox{k}\bbox{P}_{\!\!o}\over M}
                    )\, b_{\gamma\bbox{k}}
$$
and
$$S_{\text{int}}= {1\over2M}
   \sum_{\gamma\bbox{k},\gamma'\bbox{k'}}
   \int_0^t \!dt\,\bbox{k}\, b^{\star}_{\gamma\bbox{k}}  b_{\gamma\bbox{k}}
        \cdot\bbox{k'}\, b^{\star}_{\gamma'\bbox{k'}}  b_{\gamma'\bbox{k'}}
$$
A saddle point evaluation of $G_{\bbox{P}_{\!\!o}=0}(t)$  yields
$b^o_{\gamma\bbox{k}}(t)\!=
  \!-iV_{\bbox{k}}\int^{t}_0\! dt^\star e^{i(E_\gamma+k^2/2M)(t^{\star}-t)}$
and the action at the saddle point is:
$$
S^o_{\bbox{P}_{\!\!o}=0}=-t\int\!\! d\omega{\varrho_d (\omega)\over \omega}
         +i\int\!\! d\omega{\varrho_d (\omega)\over \omega^2}(e^{-i\omega t}-1)
$$
with
$$\begin{array}{ll}
 \varrho_d(\omega) & =\sum_{\gamma\bbox{k}} |V_{\bbox{k}}|^2 \,
                                      \delta(\omega-E_\gamma-{k^2\over2M}) \\
                               &=\int
d\Omega_{\hat{\bbox{k}}}\int_0^{\sqrt{2M\omega}}

dkk^{d-1}|V_k|^2\,(\omega\!-\!{k^2\over2M})
\end{array}$$
First, we consider the case of infinite mass in any dimension:
we find correctly
$\varrho_d^{\infty}(\omega)\!=\!\omega{\cal
F}(R\!=\!0)\!=\!\omega({\delta\over\pi})^2$,
i.e.\ the exponent  for the algebraic long-time decay of the heavy particle
propagator, $G(t)\!\sim\!\theta(t)\,|t|^{-\alpha}\!\!$,
  is $\alpha\!=\!({\delta\over\pi})^2$ \cite{nozieres}.

In $d=1$, ${\cal F}(R)$ is a periodic function.  Therefore its Fourier spectrum
is
discrete and only the $k=0$ component contributes to the infrared behavior so
that
$$\varrho_{\,1}(\omega)\!=\!\omega\int_0^1 d(k_f R)\,{\cal F}(R)
\quad\text{for}\quad \omega\rightarrow 0,\,M\neq\infty
$$
The OC-exponent for the algebraic long-time decay is
$\alpha\!=\!\int_0^1 d(k_f R)\,{\cal F}(R)$, i.e.\
$\alpha\!=\!f(\delta)\!\cdot\!({\delta\over\pi})^2$
with $f(\delta\!\rightarrow\!{\pi\over2})\!=\!2/3$\cite{infU} and
$f(\delta\!\rightarrow\!0)\!=\!1/2$.
It is well understood that $\alpha$ for
$M\!\rightarrow\!\infty$ is different from $\alpha$ for $M\!=\!\infty$
\cite{castella}:
the finite recoil for $M\!\neq\!\infty$ excludes scattering processes across
the Fermi sea for
$\omega\!<\!E_{\text{recoil}}$, whereas these processes cost no energy for
$M\!=\!\infty$.

Half-filling in $d=1$ is special for a
nearest neighbor hopping motion of the heavy particle. The oscillations of
${\cal F}(R)$ are commensurate with the lattice periodicity which implies
that a hopping heavy particle will probe ${\cal F}(R)$ only at its maximum
and minimum, whereas a `free' heavy particle averages over all values of
${\cal F}(R)$. Therefore we expect for this special commensurate case
$\alpha\!=\!{1\over2}(\delta/\pi)^2$, even for $\delta\!\rightarrow\!
{\pi\over2}$. This compares well with numerical findings by H.~Castella
\cite{castella}.

In $d=2$
$$\varrho_{\,2}(\omega)\!={2\over3}{\delta\tan\delta\over\pi^2}
{\omega^{3/2}\over\sqrt{k_f^2/2M}}
\quad\text{for}\quad \omega\rightarrow 0,\,M\neq\infty
$$
i.e.\ the power in $\omega$ is larger than 1 so that the QP-peak
is finite and  the particle displays coherent
behavior. The incoherent part of the spectral function diverges
for $\omega\rightarrow0$ as $1/\sqrt{\omega}$,
as already known from second order perturbation theory \cite{kopp}.
The actual value of the QP-weight and the prefactor of the incoherent part are
determined by the details of the high frequency behavior.

Only if the Fermi energy is located at a van Hove singularity
$\varrho_{\,2}(\omega)$ becomes
linear in $\omega$, as can be checked in the
2$^{\text{nd}}$-order perturbation theory. The QP-peak in
$2d$ and its suppression at a van Hove singularity was also
found numerically by S.\ Sorella \cite{sorella} using a
variational ansatz \cite{edwards}.

For $d=3$
$$\varrho_{\,3}(\omega)\!={1\over4}{\delta\tan\delta\over\pi^2}
{\omega^2\over k_f^2/2M}
\quad\text{for}\quad \omega\rightarrow 0,\,M\neq\infty
$$
which yields a QP-peak as in $d=2$, but on top of a finite incoherent
background.

We further want to mention that the step function approximation to ${\cal
F}(R)$
will produce a finite QP-peak in any dimension:
$$\varrho_{\text{step}\,d\,}(\omega)\sim
      ({\delta\over\pi})^2 M^{d/2} \omega^{d+2\over2}$$
The incorrect frequency dependence of $\varrho_{\text{step}}$ illustrates that
the
details of the long distance behavior of ${\cal F}(R)$ are indeed essential.
This result has to be contrasted with that of the non-interacting blip
approximation
which is derived by skipping the term $k^2/2M$ in $S_o$: the QP-weight is zero
in any dimension due to the missing recoil.

To estimate corrections to the saddle point result we calculated the
fluctuations at
the stationary point and the leading order shift of the saddle point by
fluctuations.
All corrections are of $O(1/M)$. Therefore, the question may be raised if the
result
for the $1d$ spectral function is exact for large $M$ and small $\omega$
($\omega\ll E_{\text{recoil}}\ll \epsilon_{k_f}$).
We expect the result for $d\!=\!1$ to be quantitatively correct
if the limit of heavy mass can be implemented by a singly retarded
action. The spectral function in $d\!>\!1$ can be
only qualitatively valid since the (non-zero) value of the QP-weight
and the prefactor of the incoherent background depend on the details
of the high energy behavior.

We acknowledge many helpful discussions with \hbox{N.~Andrei}, H.~Castella,
F.~Mila,
E.~M\"uller-Hartmann, N.V.~Prokof'ev, A.E.~Ruckenstein, and P.~W\"olfle.
This work was supported  by the DFG (T.K.).

\begin{figure}
  \caption{
   ${\cal F}(R)$ for $d=1,2,$ and $3$ with $\delta$ close to $\pi/2$.}
 \label{fig1}
\end{figure}

\begin{references}
\bibitem{highT}$R(\tau)\!=\,$const is the high-$T$ saddle point.
\end{references}
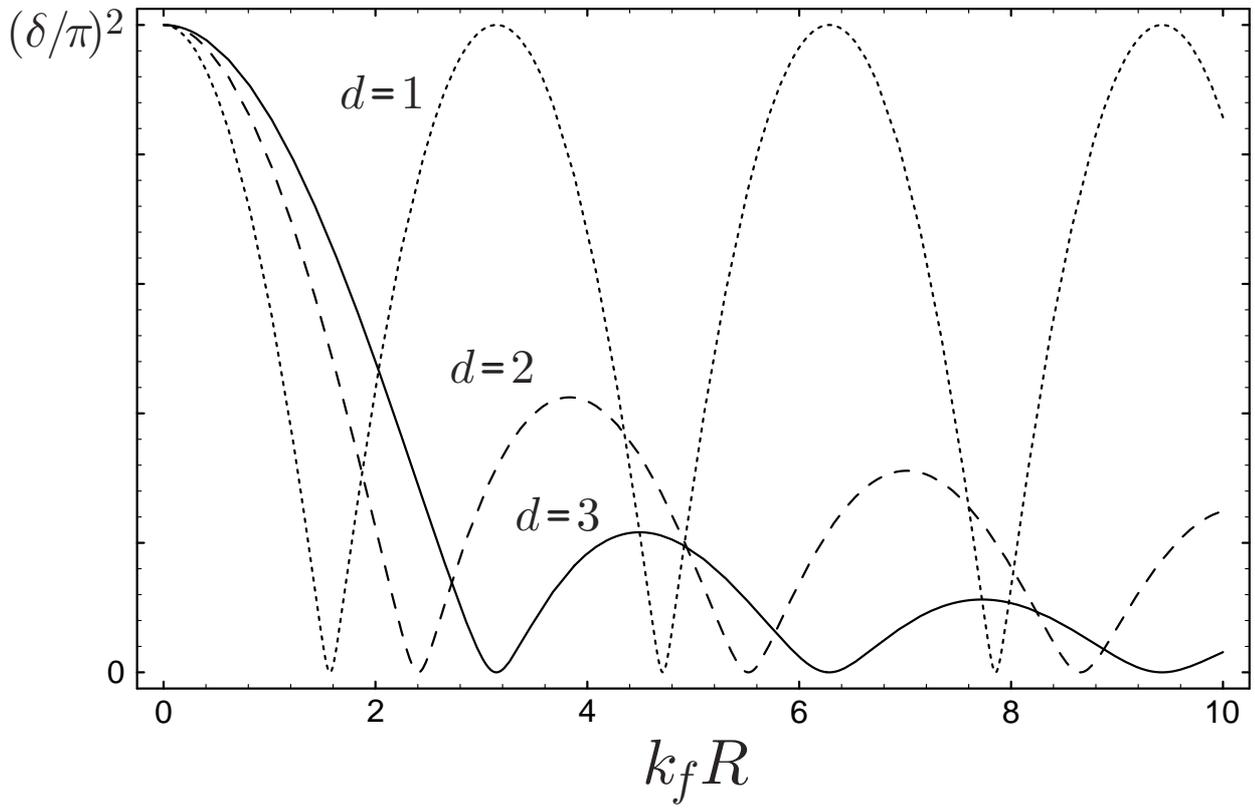
\end{document}